\documentclass[draftcls,12pt,onecolumn]{IEEEtran}
\PassOptionsToPackage{bookmarks={false}}{hyperref}
\usepackage{amssymb}
\usepackage{cite}
\usepackage{url}
\usepackage{xcolor}
\usepackage{amsmath}
\usepackage{graphicx}
\usepackage{derivative}
\usepackage{algorithm}
\usepackage{algpseudocode}
\usepackage{mathtools}
\usepackage{balance}
\usepackage{commath}
\usepackage{epstopdf}
\usepackage{blkarray} %
\usepackage{adjustbox}
\usepackage{bm}
\usepackage[hidelinks]{hyperref}

\begin{document}

	\title{Resilient and Efficient  Aerial Wireless Sensor Networks Using Pinching-Antenna Systems
	}
		\author{Ahmed A. Al-habob,~\IEEEmembership{Senior Member,~IEEE,} 
		Octavia A. Dobre,~\IEEEmembership{Fellow,~IEEE,}   
		Yindi Jing, and Hyundong Shin~\IEEEmembership{Fellow,~IEEE}    
 }

	\maketitle
	
\IEEEpeerreviewmaketitle
\begin{abstract}
This paper proposes a pinching antenna system to enhance  the connectivity and efficiency in  aerial wireless networks.  
The algebraic connectivity of the network's graph is maximized by optimizing the placement of the pinching antenna elements in the uplink and downlink waveguides. The number of deployed unmanned aerial vehicles (UAVs) is minimized while maintaining a minimum network connectivity indicator. A heuristic pinching antenna placement algorithm is designed to maximize the network connectivity. An iterative UAV deployment algorithm is also designed to find the minimum number of UAVs, their placement locations, and the pinching antenna placement. Simulation results show that the considered pinching antenna system improves network connectivity and efficiency.
\end{abstract}
\vspace{-.1in}

 \section{Introduction}
 Flexible-antenna systems are considered an enabling technology for future wireless networks. Pinching antenna systems represent a major candidate for flexible antenna designs, in which pinching elements are coupled to a dielectric waveguide to radiate radio waves at selected points along the waveguide  \cite{11222687}. This waveguide-based wave radiation architecture enables the base station to focus the  signal towards the intended users and provides spatial flexibility in optimizing the elements' placement, yielding a better link quality and enhancing
 robustness against blockage\cite{11434944}.  Pinching antenna systems have shown their potentials in  improving both uplink and downlink transmission in single-user  and multi-user settings \cite{11348983,11414143}. Furthermore, pinching antenna systems establish a strong line-of-sight link and reduce free-space path loss with aerial devices, i.e., unmanned aerial vehicles (UAVs)  \cite{11501804, 11288076, 11475785}. Different pinching antenna-enabled UAV network scenarios can be found in the literature including UAV-enabled covert communication against cooperative detection \cite{11501804},    integrated sensing and communication  \cite{11288076}, and  UAV delivery framework  \cite{11475785}. The applicability of pinching antenna systems in UAV delivery networks raises the question of their use in UAV-enabled wireless sensor networks, where power-limited devices (sensors) need to maintain  communication with UAVs.
 
 Network connectivity is an important metric in aerial wireless sensor networks, reflecting each sensor's ability to establish   communication links and transmit its sensed data to destination UAVs, either directly or via relays \cite{9994654}.  The more communication links between the sensors and the destination UAVs, the more resilient the network, enabling it to withstand unexpected disruptions such as link failures, sensor battery depletion, or cyberattacks. 
 A wireless sensor network can be represented as a graph, where nodes are vertices and communication links are edges. The network connectivity can be represented using a metric from graph theory, namely \textit{Fiedler value}, which is the second-smallest eigenvalue of the graph’s Laplacian matrix \cite{boyd2007convex}.   A non-zero Fiedler value indicates that the network is fully connected, and that each sensor can reach at least one UAV. Moreover, a higher  \textit{Fiedler value} implies more robustness and network resilience \cite{ghosh2006growing}. Some recent works have considered the \textit{Fiedler value} to study the  connectivity of  multi-modal UAV  networks \cite{11152341} and in reconfigurable intelligent surface-assisted UAV scenario  \cite{10854532}. 
 
  This paper considers a pinching antenna-aided aerial wireless sensor network, in which a base station  equipped with multiple uplink and downlink waveguides is deployed to improve the connectivity and efficiency in a multi-UAV  wireless sensor network.  The main contribution of this paper can be summarized as follows:
\begin{itemize}
	\item[$\bullet$]  A resilient multi-UAV wireless sensor network framework is developed, in which an optimization problem is formulated to maximize the \textit{Fiedler value} of the network by optimizing the placement of pinching antennas, subject to  constraints on inter-element spacing.    
	\item   To enable efficient UAV deployment while maintaining a connected network, an optimization problem is formulated to minimize the number of deployed UAVs, subject to a minimum \textit{Fiedler value} requirement and practical constraints on the UAVs placement and the pinching antennas inter-element spacing.    
	\item[$\bullet$]   A solution algorithm is developed to maximize the \textit{Fiedler value} of the network by heuristically placing the pinching elements in the uplink and downlink waveguides.   An efficient UAV placement algorithm  is also developed to determine the minimum number of required UAVs and their deployment locations.
\end{itemize}
\begin{figure}[h!]
	\centering
	\includegraphics[width=0.61\textwidth]{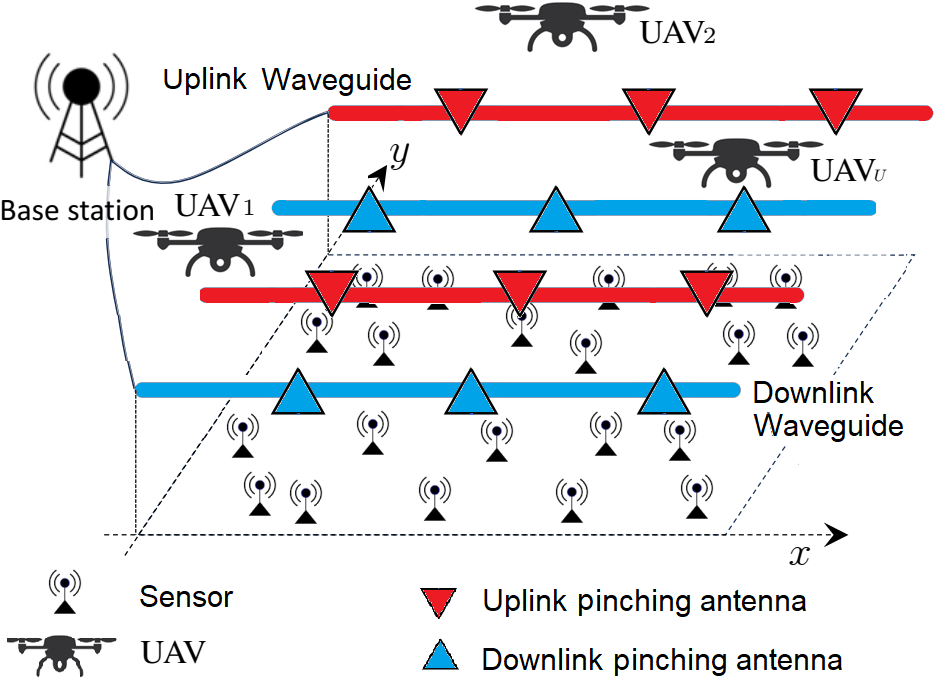}
	\caption{Aerial wireless sensor network with multi-uplink/downlink waveguides pinching antenna system.}
	\label{Sys_Mode}
\end{figure}
The remainder of this paper is organized as follows.
Section \ref{Sys}  introduces the considered aerial  sensor network, the communication channels model, and the network connectivity metric. Section \ref{ProbFormulation} discusses the formulated optimization problems.  Section \ref{Sol2} presents the developed solution algorithms and their computational complexity.     
Simulation
results are illustrated in   Section \ref{SimSec}, while Section \ref{CO} concludes the paper.

\section{System Model}\label{Sys}
As illustrated in Fig. \ref{Sys_Mode}, the considered system model consists of a base station equipped with $M$ uplink and $\bar{M}$ downlink pinched dielectric waveguides, deployed to improve connectivity in an aerial sensor network. The aerial sensor network consists of a set  $\mathcal{S} = \{s_i\}_{i=1}^{S}$ of $S$ sensors that needs to  transmit data to a set    of $U$ UAVs. Direct communication between  sensors and  UAVs can be established, which can be represented by an initial graph $\mathcal{G}'\triangleq\{\mathcal{V}', \mathcal{E}'\}$, where $\mathcal{V}'$ is a  set of $V=S+U$ vertices and $\mathcal{E}'$ is a set of   edges that represents the direct communication between the sensors and UAVs. Sensors can also establish communication links with the UAVs via the base station using the uplink and downlink pinching waveguides.  Let $\mathbf{x}_m = [x_{1,m} \cdots, x_{K_{m},m}]$ be the placement decision variable of the pinching antennas   on the $m$-th uplink   waveguide, where $K_m$ is the number of pinching antennas  in the $m$-th uplink waveguide. 
Similarly, let $\mathbf{\bar{x}}_{\bar{m}} = [\bar{x}_{1,\bar{m}} \cdots, \bar{x}_{\bar{K}_{\bar{m}},\bar{m}}]$  be the placement decision variable of the elements on the $\bar{m}$-th downlink  pinching waveguide.
 Operating the base station with the pinching antennas   adds communication links between the sensors and data gathering nodes.  
 Let $\mathcal{E}(\mathbf{X},\mathbf{\bar{X}})$ be the set of the added edges via the base station, where $\mathbf{X}=[\mathbf{x}_1,\cdots, \mathbf{x}_M] $ and $\mathbf{\bar{X}}=[\mathbf{\bar{x}}_1,\cdots, \mathbf{\bar{x}}_{\bar{M}}] $.
 The resulted graph is $\mathcal{G}(\mathbf{X},\mathbf{\bar{X}})\triangleq\{\mathcal{V}, \mathcal{E}\}$, where $\mathcal{V}=\mathcal{V}'$ and $\mathcal{E}=\mathcal{E}'\cup\mathcal{E}(\mathbf{X},\mathbf{\bar{X}})$.

\subsection{Communication Channels Model} \label{commChannel}

The  radiation vector of the $m$-th uplink and the $\bar{m}$-th downlink  waveguides can   be respectively represented
as 
\begin{equation}
	\begin{split}
		&\mathbf{r}_m\!=\!\frac{1}{\sqrt{K_m}}[\exp(-\mathfrak{j}\theta_{1,m}), \cdots, \exp(-\mathfrak{j}\theta_{K_{m},m})],\\
		&\mathbf{\bar{r}}_{\bar{m}}\!=\!\frac{1}{\sqrt{\bar{K}_{\bar{m}}}}[\exp(-\mathfrak{j}\bar{\theta}_{1,\bar{m}}), \cdots, \exp(-\mathfrak{j}\bar{\theta}_{\bar{K}_{\bar{m}},\bar{m}})],
	\end{split}
\end{equation} 
where   $\mathfrak{j} =\sqrt{-1}$ is the imaginary unit, and  $\theta_{k,m}$ and $\bar{\theta}_{k,\bar{m}}$ denote the phase shift at the $k$-th  antenna of the $m$-th uplink and $\bar{m}$-th downlink waveguides, respectively.
Let $\mathbf{g}_{i}(\mathbf{x}_m) \in \mathbb{C}^{K_{m}} =[g_{i,1}(\mathbf{x}_m), \cdots, g_{i,K_{m}}(\mathbf{x}_m)]$ be the vector of channel 
from the $i$-th sensor to pinching elements on the $m$-th uplink waveguide, such that
\begin{equation}
	g_{i,k}(\mathbf{x}_m)=\zeta\alpha_i\frac{\exp(-\mathfrak{j}\kappa \delta_{i,k}(\mathbf{x}_m) )}{\delta_{i,k}(\mathbf{x}_m)},
\end{equation}   
where $\zeta$ is a coefficient proportional to the pinching antenna's effective
surface, $\alpha_i$ represents  the shadowing coefficient, $\kappa$ is the wave number,  and  $\delta_{i,k}(\mathbf{x}_m)$ is the   distance between the $i$-th sensor and the $k$-th pinching element on the $m$-th waveguide.
The effective channel between the $i$-th sensor and the  uplink waveguides can be written as $ \mathbf{h}_i(\mathbf{X}) =[h_{i}(\mathbf{x}_1), \cdots, h_{i}(\mathbf{x}_M)]$, where $ {h}_i(\mathbf{x}_m) \triangleq  \mathbf{g}_i^T(\mathbf{x}_m) \mathbf{r}_m$, such that
\begin{equation}
	h_{i}(\mathbf{x}_m) = \zeta\alpha_i\sum_{k=1}^{K_m}\frac{\exp(-\mathfrak{j}(\kappa \delta_{i,k}(\mathbf{x}_m) +\theta_{k,m} ))}{\sqrt{K_{m}}\delta_{i,k}(\mathbf{x}_m)}.
\end{equation}

Let $\mathbf{\bar{g}}_j(\mathbf{\bar{x}}_{\bar{m}}) \in \mathbb{C}^{\bar{K}_{\bar{m}}} =[\bar{g}_{1,j}(\mathbf{\bar{x}}_{\bar{m}}), \cdots, \bar{g}_{\bar{K}_{\bar{m}},j}(\mathbf{\bar{x}}_{\bar{m}})]$ be the   channel 
from the   $\bar{m}$-th downlink waveguide to the $j$-th UAV, such that
\begin{equation}
	\bar{g}_{k,j}(\mathbf{\bar{x}}_{\bar{m}})={\zeta}	{\alpha}_j\frac{\exp(-\mathfrak{j}{\kappa} \bar{\delta}_{k,j}(\mathbf{\bar{x}}_{\bar{m}}) )}{\bar{\delta}_{k,j}(\mathbf{\bar{x}}_{\bar{m}})},
\end{equation}   
where  $\bar{\delta}_{k,j}(\mathbf{\bar{x}}_{\bar{m}})$ is the   distance between      the $\bar{k}$-th pinching element on the  $\bar{m}$-th waveguide  and the  $j$-th UAV.
The effective channel between the downlink waveguides and the  $j$-th UAV can be written as $ \mathbf{\bar{h}}_j(\mathbf{\bar{X}}) =[\bar{h}_{j}(\mathbf{\bar{x}}_{1}), \cdots, \bar{h}_{j}(\mathbf{\bar{x}}_{\bar{M}})]$,  where $ \bar{h}_{j}(\mathbf{\bar{x}}_{\bar{m}})=\mathbf{\bar{g}}^T_{k,j}(\mathbf{\bar{x}}_{\bar{m}})\mathbf{\bar{r}}_{\bar{m}}$ such that
\begin{equation}
	\bar{h}_j(\mathbf{\bar{x}}_{\bar{m}}) = \bar{\zeta}	\bar{\alpha}_j \sum_{k=1}^{\bar{K}_{\bar{m}}} \frac{\exp(-\mathfrak{j}(\bar{\kappa} \bar{\delta}_{k,j}(\mathbf{\bar{x}}_{\bar{m}}) +\bar{\theta}_{k,\bar{m}} ))}{\sqrt{\bar{K}_{\bar{m}}}\bar{\delta}_{k,j}(\mathbf{\bar{x}}_{\bar{m}})}.
\end{equation}
An orthogonal transmission scenario is considered, in which each uplink/downlink transmission takes place over orthogonal channels, whether in frequency  or time domain. The uplink  signal-to-noise ratio (SNR) at the base station
for the $i$-th sensor is defined as

\begin{equation}\label{up_SNR}
	\begin{split}
		\Gamma_i(\mathbf{X})=  \frac{P_i\abs{\sum\limits_{m=1}^{M}h_{i}(\mathbf{x}_m) }^2}{ \sigma^2},
	\end{split}
\end{equation} 
where $P_i$ is the transmit power of the $i$-th sensor and $\sigma^2$ is the power of the additive noise at the base station. The downlink SNR at     the $j$-th UAV can be written  as
\begin{equation}\label{down_SNR}
	\begin{split}
		\bar{	\Gamma}_j(\mathbf{\bar{X}})=  \frac{\abs{\sum\limits_{\bar{m}=1}^{\bar{M}}\bar{P}_{\bar{m}}\bar{h}_j(\mathbf{\bar{x}}_{\bar{m}})}^2}{ \sigma^2_j}, 
	\end{split}
\end{equation}

where $\bar{P}_{\bar{m}}$ is the transmit power of the $\bar{m}$-th waveguide and $\sigma^2_j$ is the power of the additive noise at  the $j$-th UAV.

An edge is added to the graph between the vertex representing the $i$-th sensor and the vertex representing the $j$-th UAV  if both $\Gamma_i(\mathbf{X})\geq \gamma_u$  and  $\bar{	\Gamma}_j(\mathbf{\bar{X}})\geq \gamma_d$, where $\gamma_u$ and $\gamma_d$ represent the minimum  SNR to establish reliable communication with the base station and  a UAV, respectively.

\subsection{Network Connectivity Metric}
For a given graph $\mathcal{G}$, the Laplacian matrix can be expressed as $\mathbf{L}\triangleq \mathbf{D}-\mathbf{A}$,
where $\mathbf{D}$ is the   degree matrix and $\mathbf{A}$ is the  adjacency matrix. It is worth mentioning that the  Laplacian matrix is real, symmetric and positive semi-definite, and its smallest eigenvalue $\mathcal{L}_1$ equals zero, which corresponds to the all ones eigenvector \cite{ghosh2006growing}.  
The second smallest eigenvalue $\mathcal{L}_2$ (also referred to as the \textit{Fiedler value})   represents the algebraic connectivity of the graph, which is non-zero if the graph is
connected and monotonically increasing in the edge set \cite{ghosh2006growing,boyd2007convex}, such that
\begin{equation}
	\begin{split}
&\mbox{If}~ \mathcal{G}_1\triangleq\{\mathcal{V}, \mathcal{E}_1\}, \mathcal{G}_2\triangleq\{\mathcal{V}, \mathcal{E}_2\} ~\mbox{and}~  \mathcal{E}_1\subseteq  \mathcal{E}_2,\\
&\mbox{then}~  \mathcal{L}_2(\mathcal{G}_1) \leq \mathcal{L}_2(\mathcal{G}_2). 
	\end{split}
\end{equation} 
 \section{Problem Statement}\label{ProbFormulation}
 This section introduces the problem statement of designing  a pinching antenna framework for resilient and efficient  aerial wireless sensor network.
\subsection{Resilient Aerial Wireless Sensor Networks} 
The aim is to maximize  the network connectivity and maintain a resilient network, and thus, the objective function is maximizing the \textit{Fiedler value}    of the network's graph   $\mathcal{L}_2(\mathbf{X},\mathbf{\bar{X}})$  by optimizing the uplink/downlink pinching antennas placement as follows: 
\begin{subequations}\label{P1}
	\begin{alignat}{2}
	\!\!\!	\mbox{\textbf{P1}}~	&\!\!\max_{\substack{\mathbf{X},\mathbf{\bar{X}}}}&&\mathcal{L}_2(\mathbf{X},\mathbf{\bar{X}}), \label{OP1}\\
	\!	&\mbox{s.t.}&& x_{k+1,m}-x_{k,m} \geq \Delta, \forall~m, k=1, \cdots  K_m\!-\!1,
		\label{con1}\\
		&&~~   &  \bar{x}_{k+1,\bar{m}}-\bar{x}_{k,\bar{m}} \geq \bar{\Delta}, \forall~m, k=1, \cdots  \bar{K}_{\bar{m}}\!-\!1,
		\label{con11}\\
		&&~~   &  0 \leq x_{k,m} \leq X_m, ~\forall~m, k=1, \cdots  K_m,
		\label{con2}\\
		&&~~   &  0 \leq \bar{x}_{k,\bar{m}} \leq \bar{X}_{\bar{m}}, ~\forall~m, k=1, \cdots  \bar{K}_{\bar{m}}.
		\label{con22}
	\end{alignat}
\end{subequations}
Constraints \eqref{con1} and \eqref{con11} guarantee that there is a 
 minimum spacing between any two neighbouring pinching elements to prevent mutual coupling. Constraints \eqref{con2} and \eqref{con22} guarantee that the pinching elements are placed within the length of the corresponding waveguide.   
\subsection{Efficient Aerial Wireless Sensor Networks} \label{ProbFormulation2}
An efficient aerial wireless sensor network is defined as a connected network that maintains a minimum connectivity threshold by deploying a minimum number of UAVs $\mathcal{U} \leq U$. Let $\bm{\bar{\Psi}}\triangleq[\bm{\bar{\psi}}_1, \cdots, \bm{\bar{\psi}}_{\mathcal{U}}]$, where $\bm{\bar{\psi}}_j\triangleq[\bar{\psi}_{x_j},\bar{\psi}_{y_j},\bar{\psi}_{z_j}]$ is the 3D placement  of the $j$-th deployed UAV. 
The optimization problem is formulated as follows:

\begin{subequations}\label{P2}
	\begin{alignat}{2}
		\mbox{\textbf{P2}}~	&\!\min_{\substack{\mathbf{X},\mathbf{\bar{X}}\\ \mathcal{U},\bm{\bar{\Psi}} }}&& \mathcal{U}, \label{OP2}\\
		&\mbox{s.t.}&&\mathcal{L}_2(\mathbf{X},\mathbf{\bar{X}},U,\bm{\bar{\Psi}}) \geq \varsigma,
		\label{con21}\\
		 &&~~   & z^{\min} \leq \bar{\psi}_{z_j} \leq z^{\max}, 1 \leq j\leq \mathcal{U}, \label{con222}\\
		&&~~   &   \eqref{con1}, \eqref{con11}, \eqref{con2},   ~\mbox{and}~ \eqref{con22}. 
	\end{alignat}
\end{subequations}
Constraint \eqref{con21} is the network connectivity requirement, where $\varsigma > 0$ is the connectivity threshold. Constraint \eqref{con222} represents the allowable altitude limit of the deployed UAVs.

\section{Solution Approach}\label{Sol2}
In this section,  a  heuristic pinching antenna placement algorithm  and an iterative UAV deployment algorithm  are designed to solve the formulated optimization problems as follows.  The length of the $m$-th uplink waveguide $X_m$ is divided into $N_m$ equal segments and the $\bar{m}$-th downlink waveguide $\bar{X}_{\bar{m}}$ is divided into $\bar{N}_{\bar{m}}$ equal segments. A pinching antenna element can be placed at  the middle of a segment, which satisfies \eqref{con2} and \eqref{con22}. To satisfy constraint \eqref{con1} and \eqref{con11}, the length of a segment should not  be smaller than  the minimum inter-element spacing.
Consequently, the number of segments of the $m$-th uplink and $\bar{m}$-th downlink waveguides should be   $K_m\leq N_m\leq \frac{X_m}{\Delta}$ and    $\bar{K}_{\bar{m}}\leq \bar{N}_{\bar{m}}\leq \frac{\bar{X}_{\bar{m}}}{\bar{\Delta}}$, respectively. Let   $\bm{\mathcal{X}}=[\bm{\chi}_1,\cdots, \bm{\chi}_M]$ and
$\bm{\bar{\mathcal{X}}}=[\bm{\bar{\chi}}_1,\cdots, \bm{\bar{\chi}}_{\bar{M}}]$, where $\bm{\chi}_m=[\chi_{1,m}, \cdots, \chi_{N_m,m}]$, such that $\chi_{n,m}$ (or $\bar{\chi}_{n,\bar{m}}$) is a binary variable and $\chi_{n,m} =1$ (or $\bar{\chi}_{n,\bar{m}} =1$) implies a pinching  element is placed at the center of the $n$-th segment. 

\textbf{Algorithm} \ref{euclid1} illustrates the connectivity-aware pinching antenna placement to solve \textbf{P1}, which begins by calculating the centers of the segments of the waveguides and initializing the binary variables $\bm{\mathcal{X}}$ and $\bm{\bar{\mathcal{X}}}$ as zero vectors (lines $2-3$). The algorithm iterates as long as there is an unplaced pinching element (lines $4-18$). 
At each iteration, if there is an   unplaced pinching element on the uplink waveguides, the sensor with the lowest degree in the graph ${i^*}$ is selected   and  a pinching element is placed at the center of the nearest unoccupied segment $m^*$ by setting the corresponding $\chi_{n^*,m^*}=1$ (lines $5-9$).  If there is an   unplaced pinching element on the downlink waveguides, the UAV with the lowest degree in the graph ${j^*}$ is selected  and  a pinching element is placed at the center of the nearest unoccupied segment $\bar{m}^*$ by setting the corresponding $\bar{\chi}_{\bar{n}^*,\bar{m}^*}=1$ (lines $10-15$). It is worth noting that all the elements of a waveguide could be placed while the algorithm is still iterating. Consequently,  the auxiliary variables $\omega_m$ and $\bar{\omega}_{\bar{m}}$ are introduced to remove all the segments of this waveguide from the selection process.

\begin{algorithm} 
	\caption{\small \color{black}Connectivity-aware Pinching Antenna Placement.}\label{euclid1}
	\begin{algorithmic}[1]
		\small
		\State \textbf{Input:} The initial graph $\mathcal{G}'\triangleq\{\mathcal{V}', \mathcal{E}'\}$, the locations of the sensors $\bm{\psi}_i$,   $\forall i \in \mathcal{S}$,  the locations of the UAVs $\bm{\bar{\psi}}_j$,   $ 1 \leq j \leq U$; Number of waveguide segments $N_m$ and $\bar{N}_{\bar{m}}$; 
		\State Calculate the   center of each segment in the uplink waveguides $\eta_{n,m}$ and the downlink waveguides $\bar{\eta}_{n,\bar{m}}$;
		\State  Initialize $\bm{\mathcal{X}}=\left[0 \right]_{\footnotesize{ 1\times \sum_{m=1}^{M}}N_m}$;   $\bm{\bar{\mathcal{X}}}=\left[0 \right]_{\footnotesize{ 1\times \sum_{\bar{m}=1}^{\bar{M}}N_{\bar{m}}}}$;
		\State $ ~ $\textbf{While} $\exists\sum_{n=1}^{N_m}  \chi_{n,m}<  K_m,\forall m$ \textbf{or} $\exists\sum_{n=1}^{\bar{N}_{\bar{m}}}  \bar{\chi}_{n,\bar{m}}<  \bar{K}_{\bar{m}}, \forall \bar{m}$   \textbf{do}
		{					\State $ ~~ $\textbf{If} $\exists\sum_{n=1}^{N_m}  \chi_{n,m}<  K_m,\forall m$ \textbf{do}  }
		
		\State $ ~~~ $ Select the sensor with the minimum degree in the graph:  \begin{equation*}
			i^*=\arg \min\limits_{{\forall i \in \mathcal{S}}} \{\mbox{deg}_{i}(\mathcal{E})\};
		\end{equation*}
		\State $ ~~~ $ Select the nearest unoccupied segment in the uplink waveguides: 
		\begin{equation*}
			n^*,m^*=\arg \min\limits_{\substack{1\leq n \leq N_M\\1\leq m \leq M}} \left\{\frac{\norm{\eta_{n,m}-\bm{\psi}_{i^*}}}{\omega_m[1-\chi_{n,m}]}\right\};
		\end{equation*} 
		where \begin{equation*}\label{demandind}
			\omega_m = \begin{cases} 1, &\mbox{if} ~\sum_{n=1}^{N_m}  \chi_{n,m}<  K_m,   \\ 
				0, & \mbox{otherwise}. \end{cases}
		\end{equation*} 
		\State $ ~~~ $ Set  $\chi_{n^*,m^*}=1$;
		
		{				\State $ ~~ $\textbf{End If} }
		
		{				\State $ ~~ $\textbf{If} $\exists\sum_{n=1}^{\bar{N}_{\bar{m}}}  \bar{\chi}_{n,\bar{m}}<  \bar{K}_{\bar{m}}, \forall \bar{m}$ \textbf{do} }
		\State $~~$ Select the UAV with the minimum degree in the graph:  \begin{equation*}
			j^*=\arg \min\limits_{{1 \leq j \leq U}} \{\mbox{deg}_{j}(\mathcal{E})\};
		\end{equation*}
		\State $~~$ Select the nearest unoccupied segment in the downlink waveguides: 
		\begin{equation*}
			\bar{n}^*,\bar{m}^*=\arg \min\limits_{\substack{1\leq \bar{n} \leq \bar{N}_{\bar{M}}\\1\leq \bar{m} \leq \bar{M}}} \left\{\frac{\norm{\bar{\eta}_{\bar{n},\bar{m}}-\bm{\bar{\psi}}_{{j}^*}}}{\bar{\omega}_{\bar{m}}[1-\bar{\chi}_{\bar{n},\bar{m}}]}\right\};
		\end{equation*}
		\State $~~$ where \begin{equation*}\label{demandind}
			\bar{\omega}_{\bar{m}} = \begin{cases} 1, &\mbox{if} ~\sum_{\bar{n}=1}^{\bar{N}_{\bar{m}}}  \bar{\chi}_{\bar{n},\bar{m}}<  \bar{K}_{\bar{m}},   \\ 
				0, & \mbox{otherwise}. \end{cases}
		\end{equation*} 
		\State $~~$ Set  $\bar{\chi}_{\bar{n}^*,\bar{m}^*}=1$;
		
		{						\State $ ~~ $\textbf{End If}}
		
		\State $~~$ Update the edge set $\mathcal{E}(\bm{\mathcal{X}},\bm{\bar{\mathcal{X}}})$ using \eqref{up_SNR} and \eqref{down_SNR};
		\State $~~$ Update the graph  $\mathcal{G}(\bm{\mathcal{X}},\bm{\bar{\mathcal{X}}})\triangleq\{\mathcal{V}, \mathcal{E}\}$, where $\mathcal{V}=\mathcal{V}'$ and $\mathcal{E}=\mathcal{E}'\cup\mathcal{E}(\bm{\mathcal{X}},\bm{\bar{\mathcal{X}}})$	
		\State $ ~ $\textbf{End While}
		\State \textbf{Return} $\bm{\mathcal{X}}$, $\bm{\bar{\mathcal{X}}}$, and $ \mathcal{G}\leftarrow\mathcal{G}(\bm{\mathcal{X}},\bm{\bar{\mathcal{X}}})$.
		\normalsize
	\end{algorithmic}  
\end{algorithm}

The UAV deployment approach to solve \textbf{P2} is illustrated in \textbf{Algorithm}  \ref{euclid2}, which begins by setting the number of deployed UAVs $\mathcal{U}=1$    and keeps increasing it until constraint  \eqref{con21} is satisfied (lines $2-19$). At each iteration, the sensors are clustered into $\mathcal{U}$ clusters (lines $5-16$) and a UAV is placed at the center of a cluster (lines $13-14$). Keeping in mind  constraint  \eqref{con222}, the altitude of a deployed UAV  is obtained as follows:
\begin{equation}\label{h_opt}
	\bar{\psi}_{z_j} = \begin{cases} z^{\max}, & \mbox{if }  \vartheta_{j} \tan(\theta_{\mbox{\scriptsize opt}}) \geq z^{\max}, \\ 
		z^{\min}, &  \mbox{if }  {\vartheta}_{j} \tan(\theta_{\mbox{\scriptsize opt}}) \leq z^{\min}, \\ 
		{\vartheta}_{j} \tan(\theta_{\mbox{\scriptsize opt}}), & \mbox{otherwise},\\ \end{cases}
\end{equation} 
where $\vartheta_{j}$ is the distance from the center of the $j$-th cluster to the farthest sensor in it and $ \theta_{\mbox{\scriptsize opt}}$ is a constant \cite{alzenad20173}.

\begin{algorithm} 
	\caption{\small Efficient  UAVs Deployment for Connected   Aerial Wireless Sensor Networks.}\label{euclid2}
	\begin{algorithmic}[1]
	 	\small
		\State \textbf{Input:} The locations of the sensors $\bm{\psi}_i$,   $\forall i \in \mathcal{S}$;      connectivity threshold $\varsigma > 0$; docking  locations of the available UAVs $\bm{\bar{\psi}}_j(0)$,   $\forall j=1, \cdots, U$;	maximum number of iterations $Q$;
		\State Set $\mathcal{U}=1$;  initialize the graph $\mathcal{G}'\triangleq\{\mathcal{V}', \mathcal{E}'\}$ as edgeless graph with $V=S+\mathcal{U}$; $\mathcal{G} \leftarrow \mathcal{G}'$; 		
		\State $ ~ $\textbf{While} 	$\mathcal{L}_2(\mathcal{G}) < \varsigma$ \textbf{do}		   
		 	\State $ ~~ $  Select $\mathcal{U}$ sensors and    set the initial  locations of the UAVs $\bm{\bar{\psi}}_j(1)$,   $\forall j=1, \cdots, \mathcal{U}$ to be above the selected sensors; Set  $q=1$;		   	   		
			\State $ ~~ $\textbf{While} $\norm{\bm{\bar{\psi}}_j(q)-\bm{\bar{\psi}}_j(q-1)} \geq \epsilon ~ \forall j=1, \cdots, \mathcal{U}$   \textbf{and}  $q \leq Q$ \textbf{do}
		\State $~~$  $ \bm{\mu} \leftarrow [\bm{0}]_{S\times \mathcal{U}}$;
	\State $ ~~~ $\textbf{For}  $i =1$ \textbf{to} $S$  \textbf{do}
	\State $~~~~$ $ \kappa=\arg \min\limits_{1\leq j\leq \mathcal{U}} \{\norm{{\bm{\psi}}_i - \bm{\bar{\psi}}_j(q)}\}$;
	\State $~~~~$  $\mu_{i,\kappa}=1$;
	\State $ ~~~ $\textbf{End For}
	\State $~~~$ Set $q=q+1$;
	\State $ ~~~ $ \textbf{For}  $j =1$ \textbf{to} $\mathcal{U}$  \textbf{do}
	\State $ ~~ $	  $\bar{\psi}_{x_j}(q) = \frac{\sum\limits_{i=1}^S\mu_{i,j} {\psi}_{x_i}}{\sum\limits_{i=1}^{S}\mu_{i,j}}$;    $\bar{\psi}_{y_j}(q) = \frac{\sum\limits_{i=1}^S\mu_{i,j} {\psi}_{y_i}}{\sum\limits_{i=1}^{S}\mu_{i,j}}$;  
		\State $ ~~ $ Obtain the UAV attitude $\bar{\psi}_{z_j}(q)$ using  \eqref{h_opt};
	\State $ ~~~ $ \textbf{End For}
	\State $ ~~ $\textbf{End While}
	\State $ ~ $    $\bm{\bar{\psi}}_j \leftarrow \bm{\bar{\psi}}_j(q)$, $\forall j=1, \cdots, \mathcal{U}$; Update the vertices set of the initial graph  $\mathcal{G}'$ based on $\mathcal{U}$ and  the corresponding edge set based on the SNR  threshold;
	\State $ ~ $ Obtain the pinching antenna placement $\bm{\mathcal{X}}$, $\bm{\bar{\mathcal{X}}}$ and the resulted graph $ \mathcal{G}$ using \textbf{Algorithm} \ref{euclid1};
	\State $ ~ $ $\mathcal{U}=\mathcal{U}+1$;  
	
	\State $ ~ $\textbf{End While}
	
	\State \textbf{Return} $\bm{\mathcal{X}}$, $\bm{\bar{\mathcal{X}}}$, $\mathcal{U}$, and $\bm{\bar{\psi}}_j,\forall j=1, \cdots, \mathcal{U}$.
		\normalsize
	\end{algorithmic}  
\end{algorithm}

\subsection{Computational Complexity}
Segmenting the waveguides requires $\mathcal{O}(N+\bar{N})$ operations, where $N=\sum_{m=1}^{M}N_m$ and $\bar{N}=\sum_{\bar{m}=1}^{\bar{M}} \bar{N}_{\bar{m}}$. Selecting the sensor with the minimum degree and selecting the corresponding segment require  $\mathcal{O}(S)$ and $\mathcal{O}(N)$ operations, respectively. Similarly, selecting the UAV with the minimum degree and its segment require  $\mathcal{O}(U)$ and $\mathcal{O}(\bar{N})$ operations, respectively. Updating the edge set requires $\mathcal{O}(SK+ U\bar{K})$ operations, where $K=\sum_{m=1}^{M} K_m$ and $\bar{K}=\sum_{\bar{m}=1}^{\bar{M}} \bar{K}_{\bar{m}}$. Consequently, obtaining a solution for problem \textbf{P1}  using \textbf{Algorithm} \ref{euclid1}  requires $\mathcal{O}(\tilde{K}[SK+ U\bar{K}+N+\bar{N}])$ operations, where $\tilde{K} = \max\{K,\bar{K}\}$. Clustering the sensors in \textbf{Algorithm} \ref{euclid2} requires  $\mathcal{O}(U^2SQ)$ operations and obtaining the UAV locations requires  $\mathcal{O}(U^2SQ)$ operations. Consequently,  obtaining a solution for problem \textbf{P2}  using \textbf{Algorithm} \ref{euclid2} requires $\mathcal{O}(U\tilde{K}[SK+ U\bar{K}+N+\bar{N}] +U^2SQ)$ operations.

\section{Simulation Results}\label{SimSec}
This section evaluates the performance of the proposed pinching antenna system and the designed solution algorithms. The sensors are randomly distributed in a $120 \times 20$ m rectangular area, where the uplink and downlink waveguides are alternatively equally spaced along the y-axis.   
Unless
otherwise stated, Table
\ref{TableResults} shows the parameters' numerical values.

\begin{table}[h]
	\caption{Simulation Parameters.}
	\begin{center}
		\begin{adjustbox}{width=.6\textwidth}
			\begin{tabular}{|c|c||c|c||c|c|}
				\hline
				\textbf{Parameter}&\textbf{Value} & \textbf{Parameter}&\textbf{Value}   & \textbf{Parameter}&\textbf{Value}    \\ \hline
				$ S, U$ 	&   $ 15, 5  $     &  $X_m,\bar{X}_{\bar{m}} $ 	& $ 120 $ m   &    $M, \bar{M}$    & $3$  \\ \hline  
				$ P_i$ 	&   $ 1 $ mW  &  $ K_m, \bar{K}_{\bar{m}} $	&  $ 6 $   &  $\sigma^2$, $\sigma^2_j$      &  $-90$ dBm      \\  \hline  	 	          
			\end{tabular}
		\end{adjustbox}
	\end{center}
	\label{TableResults}
\end{table}  

\begin{figure}[h!] 
	\centering
	\includegraphics[width=0.65\textwidth]{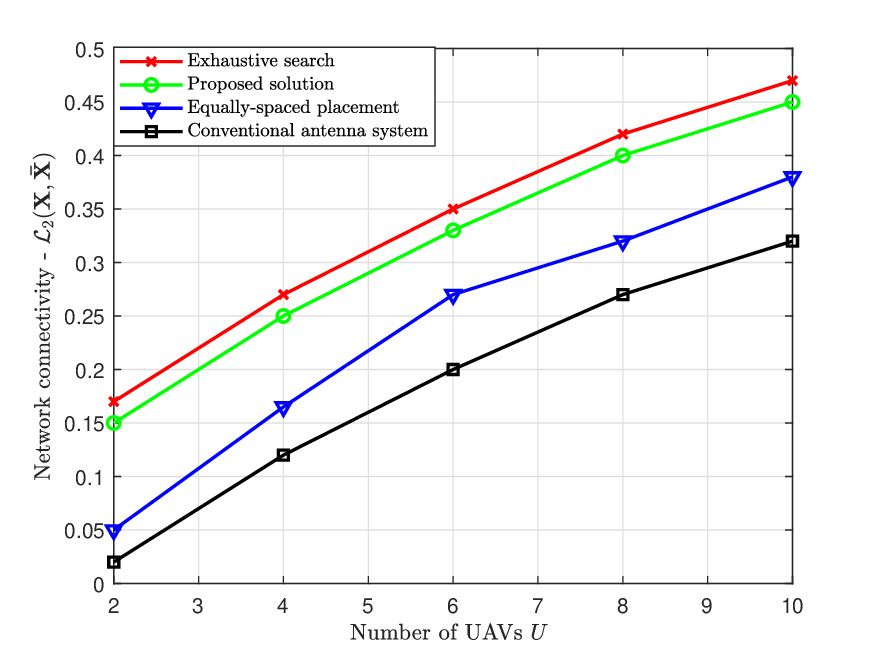}
	\caption{Network connectivity vs number of UAVs with $S=15$, $N_m=\bar{N}_{\bar{m}}=50$,  $M=\bar{M}=2$, and $K_m=\bar{K}_{\bar{m}}=4$.}
	\label{figUAV}
\end{figure}

Figure \ref{figUAV} illustrates the network connectivity vs the number of UAVs; four cases are shown: \textit{Equally-spaced placement},  in which the pinching elements are uniformly spaced in each waveguide; \textit{Exhaustive search},  in which all the possible solutions of problem \textbf{P1} are examined;  \textit{Proposed solution},    which represents the solution of \textbf{Algorithm} \ref{euclid1};  and   \textit{conventional antenna system},  in which    the base station---equipped with conventional transmit and receive antenna arrays---is located at the center of the considered area. It can be observed that the pinching antenna framework provides better connectivity in comparison with the conventional antenna system. Additionally, it can be seen that   the  proposed solution achieves better connectivity than equally-spaced placement and is close to that of the exhaustive search. Furthermore, it can be observed that deploying more UAVs enhances the network connectivity.

\begin{figure}[h!] 
	\centering
	\includegraphics[width=0.65\textwidth]{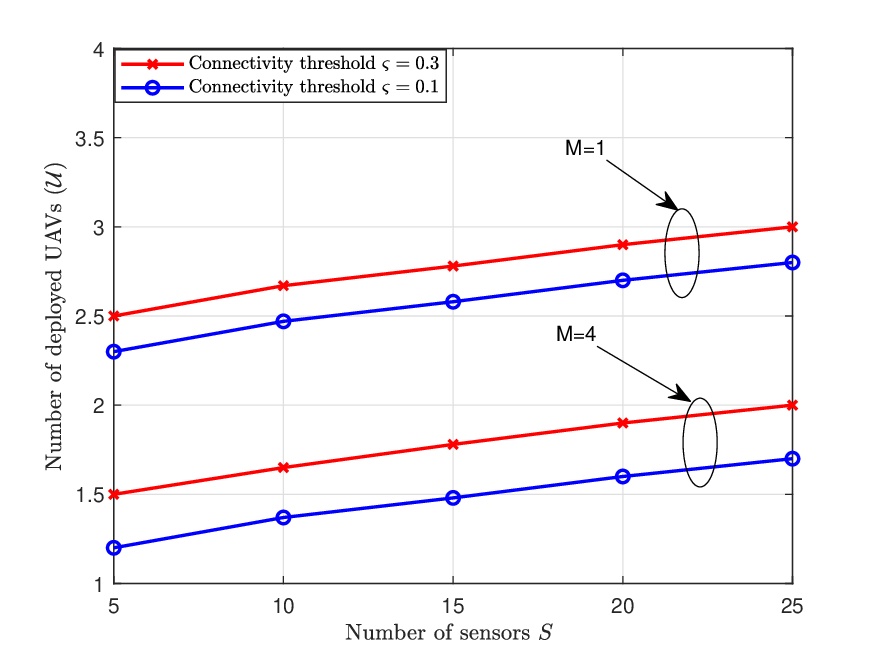}
	\caption{Number of deployed UAVs vs number of sensors with  $M=\bar{M}$,  $K_m=\bar{K}_{\bar{m}}=6$, $N_m=\bar{N}_{\bar{m}}=10^3$, and $U=7$.}
	\label{figSensors}
\end{figure}

Figure \ref{figSensors} shows the number of deployed UAVs obtained using  \textbf{Algorithm} \ref{euclid2}  vs the number of sensors, with different values of connectivity threshold $\varsigma$. It can be seen that the number of deployed UAVs increases slightly with the number of sensors. It can also be noticed that more UAVs need to be deployed as $\varsigma$ increases. Furthermore, increasing the number of waveguides and pinching elements reduces the number of required UAVs.

\section{Conclusion}\label{CO}
This paper considered a pinching antenna system to improve the connectivity and efficiency in aerial wireless networks.  
The \textit{Fiedler value}   was considered as the network resilience indicator, which was maximized by optimizing the placement of the pinching antenna elements in the uplink and downlink waveguides. The network's efficiency was studied in terms of the number of deployed UAVs required to connect the network. A heuristic pinching antenna placement algorithm was developed to maximize the network connectivity. An iterative UAV deployment algorithm was also developed to find the minimum number of UAVs, their placement locations, and the pinching antenna placement to make the network connected. Simulation results demonstrated that the considered pinching antenna system enhances the network connectivity and efficiency when compared with the conventional  system.

\bibliographystyle{IEEEtran}
\bibliography{IEEEabrv,Refrences-library}

\begin{thebibliography}{10}
\providecommand{\url}[1]{#1}
\csname url@samestyle\endcsname
\providecommand{\newblock}{\relax}
\providecommand{\bibinfo}[2]{#2}
\providecommand{\BIBentrySTDinterwordspacing}{\spaceskip=0pt\relax}
\providecommand{\BIBentryALTinterwordstretchfactor}{4}
\providecommand{\BIBentryALTinterwordspacing}{\spaceskip=\fontdimen2\font plus
\BIBentryALTinterwordstretchfactor\fontdimen3\font minus
  \fontdimen4\font\relax}
\providecommand{\BIBforeignlanguage}[2]{{%
\expandafter\ifx\csname l@#1\endcsname\relax
\typeout{** WARNING: IEEEtran.bst: No hyphenation pattern has been}%
\typeout{** loaded for the language `#1'. Using the pattern for}%
\typeout{** the default language instead.}%
\else
\language=\csname l@#1\endcsname
\fi
#2}}
\providecommand{\BIBdecl}{\relax}
\BIBdecl

\bibitem{11222687}
Z.~Yang, N.~Wang, Y.~Sun, Z.~Ding, R.~Schober, G.~K. Karagiannidis, V.~W. Wong,
  and O.~A. Dobre, ``Pinching antennas: Principles, applications and
  challenges,'' \emph{IEEE Wirel. Commun.}, vol.~33, no.~2, pp. 175--184, Apr.
  2026.

\bibitem{11434944}
Y.~Xu \emph{et~al.}, ``Generalized pinching-antenna systems: A tutorial on
  principles, design strategies, and future directions,'' \emph{IEEE Commun.
  Surv. Tutor.}, vol.~28, pp. 5872--5908, Mar. 2026.

\bibitem{11348983}
C.~Ouyang \emph{et~al.}, ``Uplink and downlink communications in segmented
  waveguide-enabled pinching-antenna systems ({SWANs}),'' \emph{IEEE Trans.
  Commun.}, vol.~74, pp. 3688--3703, Jan. 2026.

\bibitem{11414143}
A.~Bereyhi \emph{et~al.}, ``{MIMO-PASS}: Uplink and downlink transmission via
  {MIMO} pinching-antenna systems,'' \emph{IEEE Trans. Commun.}, vol.~74, pp.
  5701--5716, Feb. 2026.

\bibitem{11501804}
P.~Wu, X.~Yuan, W.~Gao, Y.~Hu, and A.~Schmeink, ``Pinching antenna assisted
  {UAV}-enabled covert communication against cooperative detection,''
  \emph{IEEE Trans. Veh. Technol.}, vol. Early Access, pp. 1--6, Apr. 2026.

\bibitem{11288076}
Y.~Hu, T.~Zhang, X.~Xu, and Y.~Liu, ``Pinching antenna-enabled integrated
  sensing and communication for low-altitude {UAV},'' \emph{IEEE Trans. Veh.
  Technol.}, vol.~75, no.~6, pp. 11\,728--11\,733, Jun. 2026.

\bibitem{11475785}
S.~Lv, M.~Li, Q.~Li, and Y.~Liu, ``Pinching-antenna system {(PASS)}-enabled
  {UAV} delivery,'' \emph{IEEE Trans. Commun.}, vol.~74, pp. 8075--8091, May
  2026.

\bibitem{9994654}
M.~Dai, N.~Huang, Y.~Wu, J.~Gao, and Z.~Su, ``Unmanned-aerial-vehicle-assisted
  wireless networks: Advancements, challenges, and solutions,'' \emph{IEEE
  Internet Things J.}, vol.~10, no.~5, pp. 4117--4147, Mar. 2023.

\bibitem{boyd2007convex}
S.~Boyd, ``Convex optimization of graph {Laplacian} eigenvalues,'' in
  \emph{Proc. International Congress of Mathematicians}, Madrid, Spain, 2007,
  pp. 1311--1319.

\bibitem{ghosh2006growing}
A.~Ghosh and S.~Boyd, ``Growing well-connected graphs,'' in \emph{Proc. IEEE
  Conference on Decision and Control}, San Diego, CA, USA, 2006, pp.
  6605--6611.

\bibitem{11152341}
Y.~Wang, J.~Farooq, and J.~Chen, ``Dynamic multi-modal {UAV} control for
  optimized coverage and backhaul connectivity in spatially unstructured and
  dispersed user environments,'' \emph{IEEE Trans. Mobile Comput.}, vol.~25,
  no.~2, pp. 2320--2334, Feb. 2026.

\bibitem{10854532}
M.~Saif and S.~Valaee, ``{RIS} alignment via virtual partitioning for resilient
  uplink multi-{RIS}-assisted {UAV} communications,'' \emph{IEEE Trans.
  Commun.}, vol.~73, no.~8, pp. 6764--6779, Aug. 2025.

\bibitem{alzenad20173}
M.~Alzenad, A.~El-Keyi, F.~Lagum, and H.~Yanikomeroglu, ``{3-D placement of an
  unmanned aerial vehicle base station ({UAV-BS}) for energy-efficient maximal
  coverage},'' \emph{IEEE Wireless Commun. Lett.}, vol.~6, no.~4, pp. 434--437,
  Aug. 2017.

\end{thebibliography}

\end{document}